# Bubble Magnetometry of Nanoparticle Heterogeneity and Interaction

A. L. Balk[1,2,3], I. Gilbert[1], R. Ivkov[4], J. Unguris[1], S. M. Stavis[1,*]

[1]Center for Nanoscale Science and Technology, National Institute of Standards and Technology, Gaithersburg, Maryland 20899, USA
[2]Maryland NanoCenter, University of Maryland, College Park, Maryland 20742, USA
[3]National High Magnetic Field Laboratory, Los Alamos National Laboratory, Los Alamos, New Mexico 87544, USA
[4]Department of Radiation Oncology and Molecular Radiation Sciences, Johns Hopkins University School of Medicine, Baltimore, Maryland 21231, USA

Bubbles have a rich history as transducers in particle-physics experiments. In a solid-state analogue, we use bubble domains in nanomagnetic films to measure magnetic nanoparticles. This technique can determine the magnetic orientation of a single nanoparticle in a fraction of a second, and generate a full hysteresis loop in a few seconds, which is much faster than any other reported technique. We achieve this unprecedented speed by tuning the nanomagnetic properties of the films, including the Dzyaloshinskii-Moriya interaction, in the first application of topological protection from the skyrmion state to a nanoparticle sensor. We demonstrate the technique on iron/nickel nanorods and iron oxide nanoparticles, which delineate a wide range of properties and applications. Bubble magnetometry enables the first measurement with high throughput for statistical analysis of the magnetic hysteresis of dispersed nanoparticles, and the first direct measurement of a transition from superparamagnetic behavior as single nanoparticles to collective behavior in nanoscale agglomerates. These results demonstrate a breakthrough capability for measuring the heterogeneity and interaction of magnetic nanoparticles.

## I. INTRODUCTION

Nanoscale manipulation is increasingly important in medicine, manufacturing, and sensing [1-3]. In environments where direct contact with a manipulator is undesirable, such as in living beings, nanoparticles enable remote manipulation [4-7]. Magnetic nanoparticles are particularly useful [8-17] due to their biological compatibility, ease of synthesis, and coupling to external fields [9,18,19]. This motivates new measurement technology, as bulk magnetometry [20,21] cannot resolve the heterogeneous properties of single nanoparticles, whereas more specialized single-particle techniques [22-28] often require meticulous preparation and are impractical for statistical analysis, which is critical for quality control and practical application [29].

Magnetic bubble domains have been demonstrated as field sensors for memory devices [30-32] and for nanoparticle magnetometry [33,34]. However, these recent measurements did not achieve significantly higher throughput or better sensitivity than other single-particle techniques. Here, we dramatically advance bubble magnetometry to measure single nanoparticles in real time, requiring only a few seconds to obtain a full hysteresis loop, which is orders of magnitude faster than other techniques [23,27]. In our technique, nanoparticles nucleate bubbles in a nanomagnetic film with perpendicular anisotropy. We expand the bubbles by applying a field of a few millitestla perpendicular to the film, and then measure the bubbles by magneto-optical Kerr effect (MOKE) microscopy. The perpendicular anisotropy of the sensor film enables simultaneous and independent modification of the magnetic state of the nanoparticle for hysteresis measurement, in contrast to techniques with higher spatial resolution [35,36].

---

[*]samuel.stavis@nist.gov

We tune both the coercive field $\mu_0H_c$ and the Dzyaloshinskii-Moriya interaction (DMI) of our films to detect millitesla fields over sensor areas of less than a square micrometer. Tuning the DMI enables the first application of topological protection from the skyrmion state of the bubbles, increasing the sensitivity and selectivity of bubble nucleation. These improvements enable, for the first time, precise statistical analysis of magnetic hysteresis loops of single dispersed nanoparticles, elucidating the propagation of heterogeneity from dimensional to magnetic property distributions. Furthermore, bubble magnetometry enables the first direct measurement of the transition in hysteresis from superparamagnetic behavior of single nanoparticles to collective behavior in agglomerates, which is relevant to cancer hyperthermia.

## II. MATERIALS AND METHODS

### A. Sensor tuning

Tuning the nanomagnetic properties of trilayer films of platinum/cobalt/platinum greatly increases their sensitivity and selectivity for transducing magnetic fields from nanoparticles into bubbles. After growth of the trilayers, we reduce their magnetic anisotropy [37], and thus bubble nucleation energy, by exposure to argon-ion irradiation. For the most sensitive measurements, we spatially vary the exposure dose [38,39] to obtain film regions near the spin-reorientation transition [40], where the film undergoes a phase transition and therefore has maximal susceptibility. To further increase sensitivity and selectivity, as we discuss below, we tune $\mu_0H_{DMI}$ to be negative, by selecting the irradiation energy to be between 80 eV and 110 eV [41].

### B. Sample nanoparticles

Iron/nickel nanorods [39,42] and iron oxide Johns Hopkins University (JHU) nanoparticles [43] delineate a wide range of relevant properties and uses. Nanorods of similar dimensions are potentially useful for magnetic actuation [44] and superconductivity [45]. JHU nanoparticles are useful for magnetic resonance contrast imaging [46] and magnetic hyperthermia for cancer therapy [47,48]. We characterize the samples by scanning electron microscopy (SEM) (Supplemental S1 and S2). The nanorods are cylindrical with lengths of 3.9 µm ± 0.5 µm and diameters of 220 nm ± 30 nm. Single JHU nanoparticles are aggregates of iron oxide crystallites, resulting in irregular shapes with approximate diameters of 100 nm ± 50 nm. Size distributions are mean values ± standard deviations. Details of sample preparation are in Supplemental S3.

### C. Signal transduction and amplification

We confirm the process of signal transduction and amplification. After deposition on the film, an anisotropic nanoparticle such as a nanorod with magnetization *M* (Fig. 1a, red cylinder) generates a fringe field *B* (Fig. 1a, blue arrows). *B* can be hundreds of millitesla at the underlying film, nucleating a bubble (Fig. 1b, white circle) near one of the magnetic poles of the nanorod with zero applied field, $B_{z(appl)}$. The pole that nucleates the bubble depends on the relative magnetization directions of the nanorod and the film. We confirm this process by scanning electron microscopy with polarization analysis (SEMPA) [36] (Fig 1b, inset). Subsequent application of $B_{z(appl)}$ expands the bubble (Fig. 1c), increasing its signal for MOKE microscopy. The center position still indicates the original nucleation position, and therefore the relative magnetization of the nanorod. The process in Fig. 1b and 1c is a single amplification cycle.

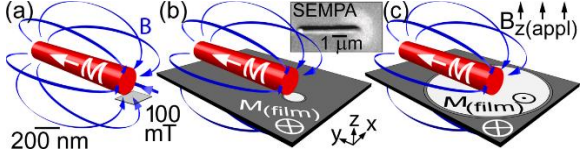

FIG. 1. Magnetic bubbles nucleate and expand underneath a magnetic nanoparticle, amplifying a magneto-optical signal. (a) An anisotropic nanoparticle such as a nanorod (red cylinder) with magnetization *M* produces a fringe field *B* (blue arrows). At the underlying plane, *B* can be hundreds of millitesla. (b) *B* nucleates a bubble (white circle) on a film with magnetization $M_{(film)}$. (Inset) Scanning electron microscopy with polarization analysis (SEMPA) shows a representative a nanorod and bubble. (c) An applied field in the *z*-direction $B_{z(appl)}$ of 5 mT expands the bubble for measurement by magneto-optical Kerr effect (MOKE) microscopy.

### D. Measurement frequencies

Many measurements per unit field are necessary to obtain hysteresis loops with high resolution. We optimize $B_{z(appl)}$ to provide hundreds of amplification cycles per second (Supplemental S4), and we measure the bubbles at a rate of 20 images per second. Aside from effects of the DMI [49], which we discuss below, in-plane magnetic fields do not influence bubble growth as the film has out-of-plane magnetization. Therefore, we can simultaneously apply an in-plane field $B_{y(appl)}$ to modify the magnetic state of a sample nanoparticle. The $B_{y(appl)}$ frequency of 50 mHz to 100 mHz is much lower than the $B_{z(appl)}$ and imaging frequencies, so the signal that we measure from the film allows readout of nanoparticle magnetization in real time.

### E. Magnetic orientation and switching

After exposure to $B_{z(appl)}$, bubbles indicating the magnetization direction of each nanorod (Fig. 2a) and some of the JHU nanoparticles (Fig. 2b) become visible in MOKE micrographs. In Fig. 2a and 2b, the magnetic moment of each sample points in the +*y* direction, in response to $B_{y(appl)} \approx$ +10 mT. JHU nanoparticles are not visible, but we infer their presence from the bubbles that nucleate and expand under them. Subsequent analysis indicates that the film senses only the largest JHU nanoparticles and agglomerates of a few JHU nanoparticles which form in liquids. Upon sweeping $B_{y(appl)}$ through zero and to -10 mT, the bubbles under the nanorod (Fig. 2c) and JHU nanoparticles (Fig. 2d) move abruptly, indicating magnetic switching of the particles. We restrict our analysis to binary magnetic switching of nanoparticles, assuming that they are single-domain structures, but it is also possible to determine multiple-domain structures.

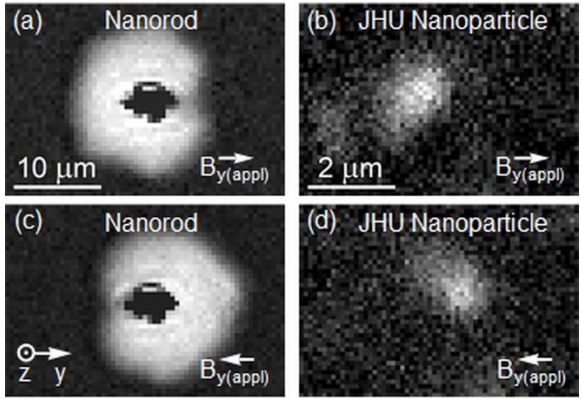

FIG. 2. MOKE micrographs showing bubble positions indicating the relative magnetization of nickel/iron nanorods and iron oxide JHU nanoparticles. (a, b) With field applied in the +*y*

direction, (a) a nanorod shows an expanded bubble (white shape) centered on its nucleation position on the left side of the nanorod (central black shape), indicating that the nanorod magnetization *M* is in the +*y* direction. (b) JHU nanoparticles are not visible but show an expanded bubble (white shape). (**c, d**) With field applied in the –*y* direction, (c) the expanded bubble moves to the right side of the nanorod, indicating that *M* has switched to the –*y* direction. (d) The expanded bubble of the JHU nanoparticles originates farther to the right than the expanded bubble in (b), indicating *M* has switched to the –*y* direction. Bright contrast indicates film magnetization in the +*z* direction. We have subtracted backgrounds from these images.

### F. Hysteresis measurements

A representative video shows magnetic switching of 14 nanorods in real time without image processing (Supplemental S5). To obtain hysteresis loops from such videos, we extract the bubble position and therefore the relative magnetization *M* of a nanoparticle by convolving each image with a kernel consisting of a positive and a negative Gaussian function on either side of the nanoparticle. Plotting the result of the convolution as a function of $B_{y(appl)}$ yields a hysteresis loop. To extract values of $\mu_0 H_c$, we fit error functions to the hysteresis loops by the method of damped least squares, quantifying uncertainties in determining the point of maximum slope.

## III. RESULTS AND DISCUSSION

### A. Measurement robustness

A series of new tests confirms that bubble magnetometry is usefully robust to various measurement parameters and film properties. We measure nanorods with a sinusoidal waveform of $B_{z(appl)}$ at a range of frequencies and amplitudes, as well as values of film $\mu_0 H_c$. The values of nanorod $\mu_0 H_c$ that we measure are independent of these parameters within uncertainty (Supplemental S6), and are also insensitive to small angles between the primary axes of the nanorods and $B_{y(appl)}$ (Supplemental S7).

### B. DMI effects

We observe that the DMI has an important effect on bubble nucleation (Supplemental S8). Briefly, if the effective DMI field $\mu_0 H_{DMI}$ is negative, then the bubbles have a domain wall chirality which matches the direction of the stray field from the nanoparticles. This reduces the nucleation energy, effectively encouraging the formation of skyrmions and increasing the sensitivity and selectivity of the film. JHU nanoparticles nucleate bubbles only on films where $\mu_0 H_{DMI}$ is negative, emphasizing the importance of controlling this property, and marking its first rational design [41] for a nanoparticle sensor. The DMI also results in asymmetric bubble expansion [49], causing a measurement artifact which we characterize in Supplemental S8.

### C. Nanoparticle hysteresis

We repeat the measurement over many cycles of $B_{y(appl)}$ to obtain a series of hysteresis loops of sample nanoparticles, elucidating behavior that would be difficult or impossible to resolve otherwise. Overlaying hysteresis loops of an exemplary nanorod (Fig. 3a) shows sharp and repeatable transitions, indicating that these anisotropic nanoparticles have exchange coupling throughout their volume. Some of the JHU nanoparticles have similar values of $\mu_0 H_c$ (Fig. 3b), however, the switching fields vary for each field cycle [50]. We observe this behavior for various excitation fields and film properties, indicating that the measurement is sensitive to stochastic switching of the nanoparticles. Further, some of the nanorods with smaller values of $\mu_0 H_c$ show similar behavior (Fig. 3c), and JHU nanoparticles with small values of $\mu_0 H_c$ have switching fields that vary by amounts approaching this value (Fig. 3d), indicating the onset of superparamagnetism. This unique capability of resolving the switching field distribution from thermal fluctuations, for many dispersed nanoparticles, allows estimation of the mean energy barrier for magnetic switching. Applying a Neel-Brown model and estimating the effective anisotropy fields $H_k$ [50,51], we calculate energy barriers of approximately 0.8 eV for the nanorods and approximately 0.1 eV for the JHU nanoparticles.

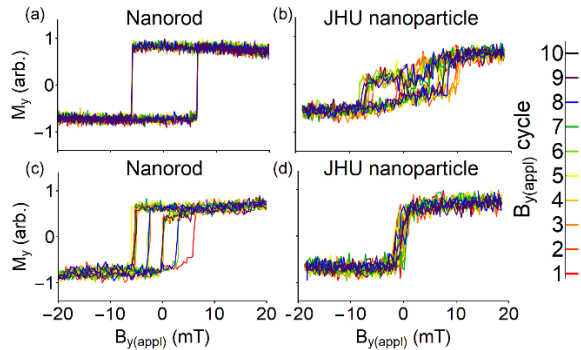

FIG. 3. Variable switching of nickel/iron nanorods and iron oxide JHU nanoparticles. (a) A nanorod with a large mean $\mu_0 H_c$ switches sharply and consistently. (b) JHU nanoparticles with similar $\mu_0 H_c$ to the nanorod in (a) switch stochastically. (c) A nanorod with a smaller mean $\mu_0 H_c$ shows several different values of $\mu_0 H_c$ from consecutive measurements. (d) JHU nanoparticles with a smaller mean $\mu_0 H_c$ show many different values of $\mu_0 H_c$.

### D. Statistical analysis

The high throughput of bubble magnetometry allows measurement of hundreds of nanorods (Fig. 4a) and JHU nanoparticles (Fig. 4b). This is an order of magnitude more than previous studies, [23,27,52] enabling precise analysis of property distributions. For the nanorods, some of the hysteresis loops are noisier due to a higher $\mu_0 H_c$ of the film, and DMI biases are evident in some of the hysteresis loops, but neither affects the $\mu_0 H_c$ values that we measure (Supplemental S6, Fig. 4). Most of the hysteresis loops have single, sharp transitions (Fig. 4a, gray circle). SEM shows that such hysteresis loops are from approximately cylindrical nanorods (Fig. 4c, left inset). A few of the hysteresis loops (Fig. 4a, black circle) show multiple, distinct switching events, from nanorods in bundles or with irregular shapes (Fig. 4c, right inset and Supplemental S1). This highlights the utility of bubble magnetometry to resolve heterogeneous magnetic properties, which are sensitive to nanoscale variation in structure. The nanorods have a mean $\mu_0 H_c$ of 7.2 mT and a standard deviation of 5.7 mT (Fig. 4c.), with a mean standard uncertainty of approximately 0.3 mT. We observe a correlation between mean $\mu_0 H_c$ and length (Supplemental S1) with an $R^2$ value of 0.8 (Fig. 4c, inset graph), which is consistent with a significant influence of shape anisotropy on magnetic properties. In comparison, the JHU nanoparticles typically have smaller $\mu_0 H_c$, with a mean of 2.2 mT and a standard deviation of 3.0 mT, with a mean standard uncertainty of 0.5 mT. Further, the smaller JHU nanoparticles all switch stochastically to some extent, due to the increasing importance of thermal effects on their magnetic properties.

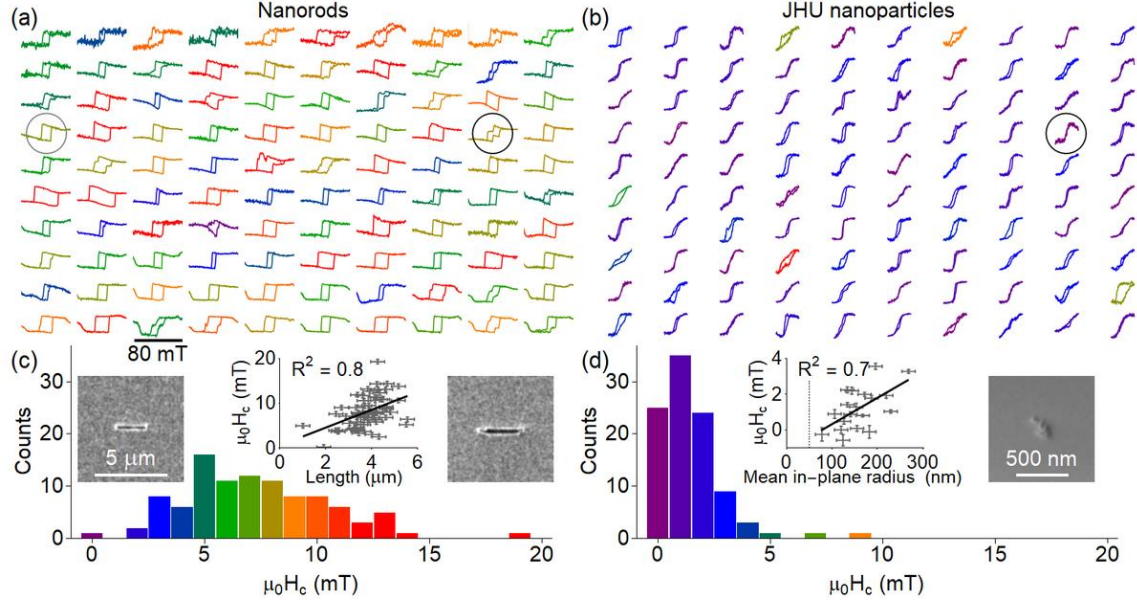

FIG. 4. Bubble magnetometry enables statistical analysis of nanoparticles. (a) Hysteresis loops of nickel/iron nanorods with colors corresponding to $\mu_0 H_c$ values. The y-axis range is arbitrary. (b) Hysteresis loops of iron oxide JHU nanoparticles, with the same $B_{y(appl)}$ and color scale as (a). (c) A $\mu_0 H_c$ histogram of the nanorods shows a mean of 7.2 mT and a standard deviation of 5.7 mT. (Inset plot) Correlation of mean $\mu_0 H_c$ and nanorod length, showing the influence of shape anisotropy. (Inset images) SEM micrographs of a cylindrical nanorod and a nanorod bundle, corresponding to the hysteresis loops circled in (a). (d) A $\mu_0 H_c$ histogram of the JHU nanoparticles shows a mean of 2.2 mT and a standard deviation of 3.0 mT. (Inset plot) Correlation of mean $\mu_0 H_c$ and mean in-plane radius of JHU nanoparticles and agglomerates. Extrapolating (solid line) to the mean radius of single JHU nanoparticles (dashed line) shows that many JHU nanoparticles have vanishing $\mu_0 H_c$, thus exhibiting superparamagnetic behavior at measurement frequencies of less than 1 Hz. Negative values of $\mu_0 H_c$ could result from stochastic switching. (Inset image) SEM micrograph of the smallest measurable JHU nanoparticle, corresponding to the hysteresis loop circled in (b). In both inset plots, vertical bars are standard uncertainties, and horizontal bars are limits of uncertainty (Supplemental S1 and S2).

### E. Nanoparticle superparamagnetism

For the JHU nanoparticles, many $\mu_0 H_c$ values approach 0 mT, so we hypothesize that some particles are superparamagnetic under our measurement conditions. To test this, we characterize a subset by SEM to determine their in-plane sizes (Supplemental S2). The results are consistent with a previous study [43], although we observe larger particles with a greater variety of shapes, indicating that this subset ranges from single JHU nanoparticles near the large end of their size distribution to agglomerates of a few JHU nanoparticles. The mean $\mu_0 H_c$ of these nanoparticles or agglomerates correlates with their mean in-plane radius (Fig. 4d inset plot), with an $R^2$ value of 0.7. Extrapolating this trend to the mean radius of single JHU nanoparticles of approximately 50 nm (Fig. 4d inset, dashed line) shows that many have vanishing $\mu_0 H_c$ and therefore are superparamagnetic at our measurement frequency of < 1 Hz. From the smallest measureable JHU nanoparticle (Fig. 4d, circled hysteresis loop, inset image), we estimate the moment sensitivity of our technique as $5 \times 10^{16}$ A·m$^2$. This compares favorably with direct Kerr magnetometry to obtain slightly better sensitivity [53], with lower throughput and the requirement of reflective samples.

**F. Agglomerate behavior**

Exchange coupling between JHU nanoparticles in an agglomerate is unlikely, and the largest particles that we measure from this sample are larger than we expect for the population [43]. Therefore, the correlation of $\mu_0 H_c$ and mean in-plane radius, for radii larger than single JHU nanoparticles, indicates that nanoparticle fringe fields mediate their collective behavior, resulting in a superferromagnetic or a superspin glass state within agglomerates [54]. In these states, single nanoparticles are superparamagnetic but dipolar interactions cause collective behavior, leading to magnetic hysteresis of agglomerates [55]. Previous studies have reported evidence for such behavior in ensemble measurements of nanoparticles in granular films [56-58], two-dimensional arrays [59,60], and quasi-two-dimensional and one-dimensional chains [61]. In comparison, bubble magnetometry enables the first direct measurement of the transition in hysteresis from superparamagnetic behavior of single nanoparticles to their collective behavior in nanoscale agglomerates. This result emphasizes the importance of isolating nanoparticle interactions, which can confound ensemble magnetometry and affect nanoparticle function in critical applications. In particular, nanoparticles commonly agglomerate in biological media [62,63] and their resulting properties strongly influence heating efficiency [64] in cancer hyperthermia. Future measurements at higher frequencies will further elucidate such structure-property relationships.

**IV. CONCLUSION**

We have developed a magnetometry technique, based on bubble domain nucleation and expansion, which generates hysteresis loops of single nanoparticles in a few seconds. This is orders of magnitude faster than other reported techniques. We have achieved this unprecedented throughput by tuning the nanomagnetic properties of our sensor films, including the first rational design of the DMI for such devices. This has enabled application of this technique to elucidate the physical properties of dispersed nanoparticles, including their heterogeneity and interaction. Bubble magnetometry can not only facilitate fundamental study of magnetic nanoparticles, but also meet the critical challenge of statistical analysis for quality control [29]. This will enable emerging technologies which rely on magnetic nanoparticles [46,48], and foster further development of technology for measurement [34,65] and application of magnetic nanoparticles.

**ACKNOWLEDGMENTS**
The authors acknowledge Emily Follansbee for assistance with figure preparation, Lamar Mair and Carlos Hangarter for assistance with sample preparation, and Mark Stiles for helpful comments. A.L.B. acknowledges support of this research under the Cooperative Research Agreement between the University of Maryland and the National Institute of Standards and Technology Center for Nanoscale Science and Technology, Award No. 70NANB10H193, through the University of Maryland, as well as the Laboratory Directed Research and Development Program at Los Alamos National Laboratory.

# Supplemental Material *for*
# Bubble Magnetometry of Nanoparticle Heterogeneity and Interaction

**CONTENTS**
S1. Scanning electron microscopy of nanorods
S2. Scanning electron microscopy of JHU nanoparticles
S3. Sample preparation
S4. $B_{z(appl)}$ excitation scheme
S5. Video of bubble magnetometry in operation
S6. Measurement robustness to excitation waveform and film properties
S7. Measurement robustness to angle between applied field and nanoparticle axis
S8. Effects of the Dzyaloshinskii-Moriya interaction (DMI)
References

**S1. Scanning electron microscopy of nanorods**

Fig. S1 shows a scanning electron micrograph of representative iron/nickel nanorods on a ferromagnetic film. Large agglomerates of nanorods are easily identifiable in both MOKE micrographs and during bubble magnetometry, and we do not consider them further. Instead, we measure single nanorods, as well as a few bundles of two nanorods. Most of the nanorods are approximately cylindrical, although a few structural variations such as forks are evident.

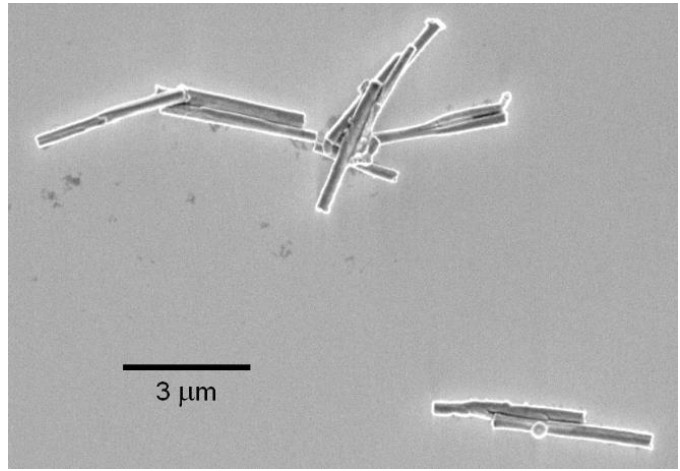

FIG. S1. Scanning electron micrograph showing representative nanorods.

We measure the in-plane length of each nanorod by visually determining the positions of the nanorod ends in the images. We estimate conservative limits of uncertainty from our ability to differentiate between the nanorod and the surrounding area in the micrograph. We assume that the in-plane length of the nanorod and its actual length are equal in this analysis.

**S2. Scanning electron microscopy of JHU nanoparticles**

Fig. S2 shows scanning electron micrographs of JHU nanoparticles on a ferromagnetic film, corresponding to the inset plot of Fig. 4d of the main text. We interpret these to be single JHU nanoparticles near the large end of their size distribution, or agglomerates of a few JHU nanoparticles. We cannot reliably distinguish between the two, due to the heterogeneous sizes and shapes of single JHU nanoparticles. Further, these nanoparticles are themselves aggregates of multiple crystallites of iron oxide with ambiguous boundaries in an agglomerate of multiple JHU nanoparticles, at least at the resolution of our scanning electron micrographs.

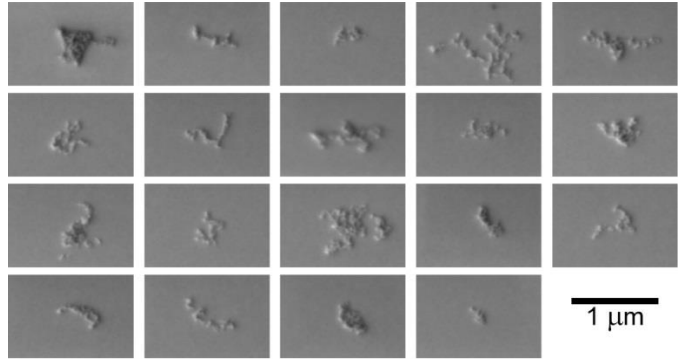

FIG. S2. Scanning electron micrographs showing the subset of JHU nanoparticles from the inset plot of Fig. 4d of the main text. The scale is the same for each image.

To measure the mean in-plane radius of a single JHU nanoparticle or agglomerate of a few JHU nanoparticles, we subtract the mean value of each image and square each pixel value, since the nanoparticles have both bright and dark contrast in the SEM images. Then, we spatially filter and binarize the images by Otsu's method. We derive the parameters of the spatial filter by visually comparing the images before and after binarization. We determine the centroid positions and mean radii of the resulting shapes. We derive limits of uncertainty from the sensitivity of the values of mean radii to the parameters of the spatial filter.

### S3. Sample preparation

We prepare nanoparticles for measurement by dispersing them in pure water, sonicating the suspension, and depositing drops of it onto the film. We deposit most nanorods on a film with $\mu_0 H_{DMI} \approx 10$ mT, some nanorods on a film with $\mu_0 H_{DMI} \approx -10$ mT for test measurements, and all JHU nanoparticles on a film with $\mu_0 H_{DMI} \approx -10$ mT to increase sensitivity. We measure all JHU nanoparticles near the spin reorientation transition of the film [1], where it is most sensitive. To align the easy axes of the nanoparticles, we apply a magnetic field of approximately 10 mT in the $y$ direction while the suspension dries. We adjust the suspension concentration so that the typical separation distance between nanoparticles after drying is tens of micrometers. Finally, we expose the JHU nanoparticles to ultraviolet ozone to further increase sensitivity (not shown), probably by removing the citric acid coating and moving the nanoparticles closer to the film.

### S4. $B_{(appl)}$ excitation scheme

We apply $B_{z(appl)}$ with a coreless electromagnet under the film. We use different $B_{z(appl)}$ excitation schemes to measure hysteresis loops of iron/nickel nanorods and iron oxide JHU nanoparticles. The nanorods show relatively strong signals, so we excite them with a sinusoidal waveform with a frequency of 1 kHz and a negative offset. The offset is much less than the amplitude, ensuring total annihilation of bubbles after each field peak. The JHU nanoparticles require a different $B_{z(appl)}$ waveform to optimize measurement of their relatively weak signals. This waveform (Fig. S4) begins with a negative field pulse to initially saturate the ferromagnetic film in a negative condition, immediately after which a positive pulse expands the bubble nucleated under the nanoparticle. The total duration of these two pulses is the time $\tau_1$. Then, $B_{z(appl)}$ returns to a negative bias value, and remains at the bias value for a longer time $\tau_2$. The negative bias field ensures initialization after each amplification cycle, but is small enough so as not to change the size of the bubbles during $\tau_2$. As $\tau_2 \gg \tau_1$, the bubbles are approximately stationary during most of the measurement, increasing their signal-to-noise ratio in MOKE micrographs. In this way, we can perform bubble magnetometry measurements at a frequency of 20 Hz. Our technique also allows measurements close to the spin reorientation transition of the film, where its spontaneous demagnetization limits the duration of each measurement cycle.

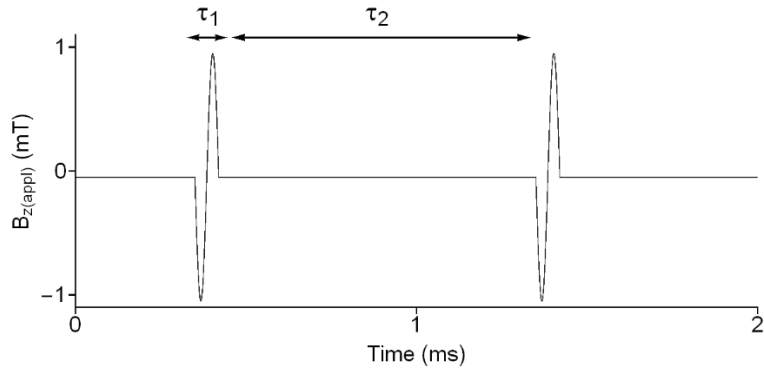

FIG. S4. A simulation showing the $B_{z(appl)}$ waveform that we use to measure JHU nanoparticles. The pulses initialize the ferromagnetic film in the -$z$ direction, allow bubble nucleation, and expand the bubble for measurement. The field returns to a small bias value and remains there until the next pulses. The total duration of the positive and negative pulses $\tau_1$ is much less than the duration of the bias value $\tau_2$, so the bubbles are nearly stationary during the measurement.

We apply $B_{y(appl)}$ with Helmholtz coils which we adjust to be coplanar to the film surface and measure $B_{y(appl)}$ with an *in situ* Hall sensor with a much higher bandwidth. We estimate the limit of misalignment between the film and $B_{y(appl)}$ to be $2\times10^{-2}$ rad, from the range of angles that lead to visible asymmetries in bubble growth due to $B_{y(appl)}$ during measurement. This misalignment would lead to an out-of-plane field of 0.5 mT, which is too small, in comparison to the stray field from the nanoparticles of tens of millitesla, to significantly affect domain nucleation.

### S5. Video of bubble magnetometry in operation

Video S5 shows bubble magnetometry in real time. In this video, nanorods scatter light, appearing as irregular shapes with white contrast. The magnetization direction of each nanorod is visible as the position of the underlying bubble, which appears as a roughly circular patch with light gray contrast. In this video, the ferromagnetic film responds to excitation by $B_{z(appl)}$ as we describe in S4 and $B_{y(appl)}$ at a frequency of 0.5 Hz and at an amplitude of 30 mT. 14 nanorods (numbered) show switching. Other bubbles are from film defects but do not show hysteresis.

### S6. Measurement robustness to excitation waveform and film properties

We measure $\mu_0 H_c$ of nanorods using variable parameters to test for artifacts. The values of $\mu_0 H_c$ that we measure are independent of the amplitude (Fig. S6a) and frequency (Fig. S6b) of the $B_{z(appl)}$ waveform. Further, since measuring on regions of the film with higher $\mu_0 H_c$ requires larger $B_{z(appl)}$, which could potentially affect the magnetic states of the nanorods, we check for correlation between the $\mu_0 H_c$ values of the 100 nanorods in Fig. 4a and the $\mu_0 H_c$ values of the ferromagnetic film proximate to each nanorod (Fig. S6c), finding these two parameters to be independent. These results are consistent with bubble nucleation occurring at faster time scales than $B_{z(appl)}$ excitation, and confirm the robustness of bubble magnetometry.

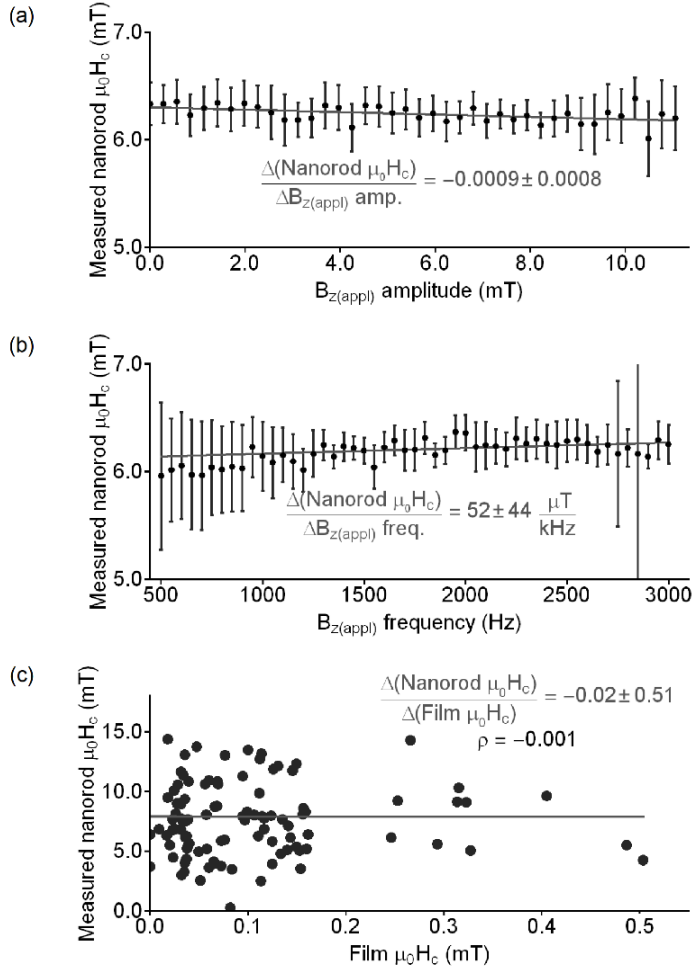

FIG. S6. Bubble magnetometry is robust against variable measurement parameters. (a) $\mu_0 H_c$ values that we measure for nanorods are independent of the amplitude of the $B_{z(appl)}$ waveform. (b) $\mu_0 H_c$ values that we measure for nanorods are independent of the frequency of the $B_{z(appl)}$ waveform. Vertical bars in (a) and (b) are standard uncertainties. (c) $\mu_0 H_c$ values that we measure for nanorods are insensitive to $\mu_0 H_c$ values of the film, which vary by more than an order of magnitude. Uncertainties in values of the $\mu_0 H_c$ of the nanorods are typically less than 0.3 mT, and uncertainties in the values of $\mu_0 H_c$ of the film are typically less than 20 $\mu$T.

Although these results show that the technique is practical on films with higher values of $\mu_0 H_c$, the highest signal-to-noise ratio results from areas of the film with lower $\mu_0 H_c$, due to lower stochastic domain wall pinning and higher domain wall mobility. This explains the lower signal-to-noise ratio of the hysteresis loops at the top of Fig. 4a of the main text.

**S7. Measurement robustness to angle between applied field and nanoparticle axis**

The high throughput of bubble magnetometry allows us to measure many hysteresis loops of the same nanorod while applying $B_{y(appl)}$ along a range of angles $\theta$ with respect to the long axis of the nanorod (Fig. S7). From these test measurements, we determine $\mu_0 H_c$ as a function of $\theta$, finding that $\mu_0 H_c$ is insensitive to $\theta$ for small $\theta$. Therefore, the limit of misalignment between the nanorods and $B_{y(appl)}$ of 0.1 rad has a negligible effect on the hysteresis loops and $\mu_0 H_c$ values in Fig. 4 of the main text.

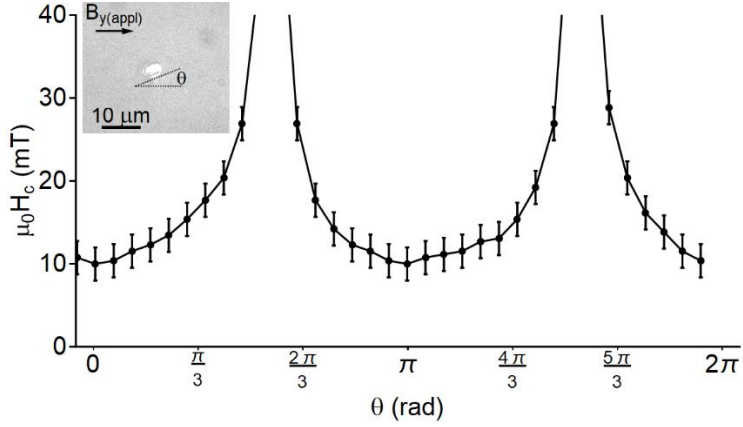

FIG. S7. A plot of $\mu_0 H_c$ as a function of the angle $\theta$ between $B_{y(appl)}$ and the long axis of a nanorod shows that $\mu_0 H_c$ values are insensitive to $\theta$ for small $\theta$. Vertical bars are standard uncertainties.

## S8. Effects of the Dzyaloshinskii-Moriya interaction (DMI)

We find that the DMI affects film sensitivity and results in measurement artifacts. We characterize the DMI as an effective in-plane field, $\mu_0 H_{DMI}$, acting on the magnetic moment $\boldsymbol{M}_{DW}$ within the domain walls surrounding the bubbles. $\mu_0 H_{DMI}$ causes $\boldsymbol{M}_{DW}$ in bubbles with moment in the $+z$ direction to point towards the center of the bubbles for regions with $\mu_0 H_{DMI} < 0$ (Fig. S8a, left, black arrows), and away from the center of the bubble for regions with $\mu_0 H_{DMI} > 0$ (Fig. S8a, right, black arrows). The Zeeman interaction between $\boldsymbol{M}_{DW}$, the applied field $B_{y(appl)}$, and the fringe field from the particles $\boldsymbol{B}$, affects both nucleation and expansion of bubbles. During bubble nucleation, $\boldsymbol{M}_{DW}$ interacts with $\boldsymbol{B}$ (blue arrows) from an anisotropic nanoparticle (red cylinder). In a film with $\mu_0 H_{DMI} < 0$, $\boldsymbol{B}$ and $\boldsymbol{M}_{DW}$ align, decreasing the energy for bubble nucleation and increasing the sensitivity of the film (Fig. S8a, left). In a film with $\mu_0 H_{DMI} > 0$, $\boldsymbol{B}$ and $\boldsymbol{M}_{DW}$ misalign, increasing the energy for bubble nucleation and reducing the sensitivity of the film (Fig. S8a, right). Therefore, in bubble magnetometry, it is advantageous for the film to have $\mu_0 H_{DMI} < 0$. To quantify this alignment effect, we calculate the nucleation energy as a function of bubble radius for typical magnetic parameters of the film in Fig. S8b, with the $\mu_0 H_{DMI} > 0$ nucleation energy in black and the $\mu_0 H_{DMI} < 0$ energy in gray. The DMI significantly affects the total nucleation energy. For this reason, the JHU nanoparticles do not nucleate bubbles on the film with $\mu_0 H_{DMI} > 0$, motivating their measurement on a film with $\mu_0 H_{DMI} < 0$.

During bubble expansion, in contrast, the applied in-plane field $B_{y(appl)}$ modifies the surface energy of the domain wall, due to Zeeman energy between $B_{y(appl)}$ and $\boldsymbol{M}_{DW}$. This changes the propagation velocity of the domain wall. Since $\boldsymbol{M}_{DW}$ on either side of the bubble has opposite directions, the DMI leads to asymmetric bubble expansion [2], biasing the hysteresis data.

To characterize these biases, we measure 10 nanorods on a film that we prepare [3] to have $\mu_0 H_{DMI} \approx 10$ mT and 10 nanorods on a film with $\mu_0 H_{DMI} \approx -10$ mT. Biases from the DMI appear in hysteresis loops as backgrounds that are not hysteretic and that depend on the applied field, with a positive derivative at $B_{y(appl)} = 0$ for regions with $\mu_0 H_{DMI} > 0$ (Fig. S8c, black loop), and a negative derivative at $B_{y(appl)} = 0$ for regions with $\mu_0 H_{DMI} < 0$ (Fig. S8c, grey loop). Such backgrounds should not affect measurements of $\mu_0 H_c$, however, which we verify by comparing the values of $\mu_0 H_c$ for the hysteresis loops in Fig. 3, which are both 6 mT ± 1 mT. For accurate measurements of hysteresis loop shape, these biases are amenable to further analysis by measuring $\mu_0 H_{DMI}$ and bubble expansion in the film without a sample nanoparticle [2].

As an example of such analysis, we determine the influence of $\mu_0 H_{DMI}$ on bubble expansion from the hysteresis loops in Fig. S8c. We do so by extracting the two portions of a hysteresis loop with decreasing absolute value of $B_{y(appl)}$, where the magnetic state of the nanorods, and therefore the bubble nucleation location, is not changing. We remove the offset between these two portions, resulting in an estimate of the difference in domain wall velocity between the right and left sides of the bubble, without the effect of the nanoparticle (Fig. S8d). These results are consistent with previous studies of domain wall motion in similar films [2,3], and allow estimation of $\mu_0 H_{DMI}$ for these two films. Subtraction of these curves from the corresponding hysteresis loops would eliminate artifacts due to the DMI.

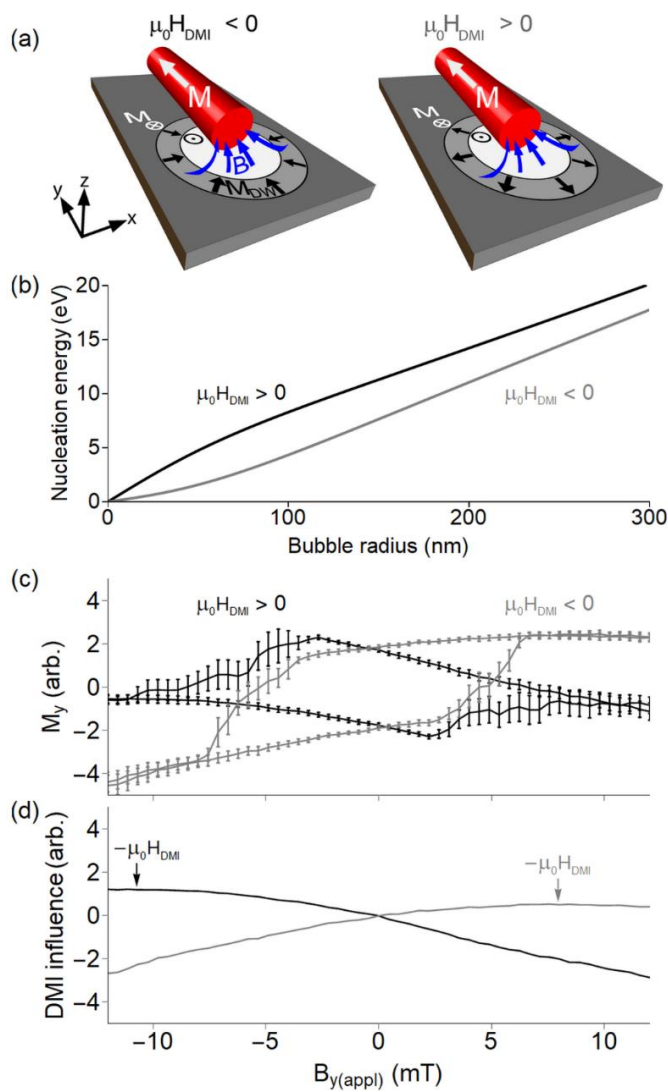

FIG. S8. DMI effects on bubble nucleation and expansion in bubble magnetometry. (a) Left: An anisotropic nanoparticle such as a nanorod (red cylinder) nucleates a bubble in the $+z$ direction (white circle) in a film with $\mu_0 H_{DMI} < 0$. The negative DMI causes the moment within the domain wall $\mathbf{M}_{DW}$ (black arrows) to point toward the center of the bubble, which is in approximately the same direction as the fringe field from the nanorod $\mathbf{B}$ (blue arrows). This near

alignment reduces the energy barrier for bubble nucleation, increasing film sensitivity. Right: A nanorod nucleates a bubble in the $+z$ direction in a film with $\mu_0 H_{DMI} > 0$. The positive DMI causes $\boldsymbol{M}_{DW}$ to point away from the center of the bubble, which is not in the same direction as $\boldsymbol{B}$. This misalignment increases the energy barrier for nucleation. (b) Calculation of the energy for creating a bubble in a film with $\mu_0 H_{DMI} > 0$ (black) and $\mu_0 H_{DMI} < 0$ (gray) as a function of the bubble radius. (c) The DMI also causes different biases in hysteresis loops of ten nanorods from regions of the film with $\mu_0 H_{DMI} < 0$ (gray data) and $\mu_0 H_{DMI} > 0$ (black data). However, these biases do not affect the values of $\mu_0 H_c$ that we extract from these two hysteresis loops, which are both 6 mT ± 1 mT. Vertical bars are standard uncertainties. (d) Data showing the DMI effect on the measurement, from the portions of the hysteresis loops in (c) with decreasing absolute value of field. Arrows indicate approximate values of $\mu_0 H_{DMI}$ for the two films.